\documentclass{edbk} 
  \usepackage{psfig}

\kluwerbib

\begin{document}

\def\sss#1{{{\scriptscriptstyle #1}}}
\def\ah{{\scriptscriptstyle 1/2}}
\def\Teff{{$T_{ef\!f} $}}
\def\Mo{{M$_\odot $}}
\def\Lo{{L$_\odot $}}
\def\apj{{ApJ~}}
\def\eg{{\it e.g.~}}
\def\cf{{\it c.f.~}}
\def\ie{{\it i.e.~}}
\def\qv{{\it q.v.~}}
\def\at{{\rm\char'100}}
\def\viz{{\it viz.~}}
\def\etal{{\it et al.~}}
\def\apriori{{\it a priori~}}
\def\loccit{{\it loc. cit.~}}
\def\Teff{{T$_{ef\!f}$}}
\def\th{\thinspace}
\def\ni{\noindent}
\def\Xi{{${\mathbf X_{i}}~$}}
\def\Xip{{${\mathbf X_{i+1}}~$}}
\def\F{{${\mathbf F}~$}}
\def\approxgt{\,\raise2pt \hbox{$>$}\kern-8pt\lower2.pt\hbox{$\sim$}\,}
\def\approxlt{\,\raise2pt \hbox{$<$}\kern-8pt\lower2.pt\hbox{$\sim$}\,}
\def\hf{\hfill}
\def\o{{\rm o}}

\booktitle[Nonlinear Stellar Pulsation]
{Nonlinear Stellar Pulsation}




\articletitle{Nonlinear Analysis of Irregular Variables}


\author{J. Robert Buchler}
\affil{Physics Department, University of Florida\\
Gainesville FL32611, USA}
\email{buchler@physics.ufl.edu}

\author{Zolt\'an Koll\'ath}
\affil{Konkoly Observatory\\
P.O. Box 67, H-1525 Budapest, Hungary}
\email{kollath@konkoly.hu}


\begin{abstract}
The Fourier spectral techniques that are common in Astronomy for analyzing
periodic or multi-periodic light-curves lose their usefulness when they are
applied to unsteady light-curves.  We review some of the novel techniques that
have been developed for analyzing irregular stellar light or radial velocity
variations, and we describe what useful physical and astronomical information
can be gained from their use.
\end{abstract}


\vskip-16cm
{\large \noindent{\bf NONLINEAR STELLAR PULSATION}\\
\noindent{\bf Editors: M. Takeuti \& D. Sasselov}\\
\noindent{\bf in Astrophysics and Space Science Library (ASSL)}

\vskip 15cm

\section{Introduction}

Is it possible, from the measurement of an irregular light-curve,
to infer something interesting and physically useful about the
mechanism that generated it?  This is the fundamental question
that we address in this review paper.

The first thing that comes to mind when confronted with a time-series is to
perform a Fourier spectral analysis.  Such an analysis is of course extremely
useful when the signal is periodic or multi-periodic, but these two types of
signals do not exhaust all possibilities.  Signals can also be  stochastic or
chaotic, in which case the spectral analysis is not particularly
insightful. \th (For a good general discussion of why the Fourier analysis or
any other linear analyses such as ARMA break down \cf Weigend \& Gershenfeld,
1994 and Abarbanel \etal 1993).

As an example and a challenge to the astronomer reader we present the
time-series shown in Fig.~\ref{roessts}.  The time-series shows large amplitude
variations, yet according to the Fourier spectrum in Fig.~\ref{roessft}, the
signal appears to be mono-periodic.  An O-C (observed-calculated) diagram
displays phase meanderings.  So what is the nature of this signal?  In
anticipation we note that this signal is the solution of a {\it deterministic}
third order ordinary differential equation.  It never repeats itself despite
the apparent mono-periodic spectrum.  In the nonlinear dynamics parlance this
signal falls into the class of low dimensional chaos.

\begin{figure}
\centerline{\psfig{figure=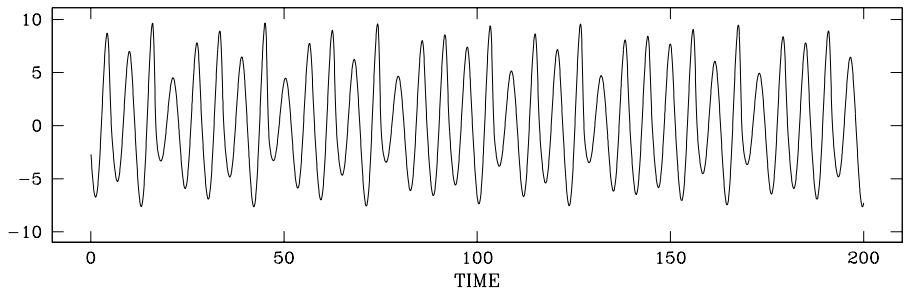,width=13.5cm}}
\caption{\noindent Temporal behavior of variable $x_1(t)$ of the
R\"ossler oscillator.}
\label{roessts}
\end{figure}

The signal that we have just discussed is a paradigm for understanding certain
quite common types of real observational signals (even though of course the
latter will in addition be contaminated by noise).  We shall now describe in
very simple terms three complementary techniques that we have found useful for
analyzing such signals, namely an explicit functional reconstruction of the
dynamics (sometimes referred to as global flow reconstruction), a topological
analysis and a time-frequency analysis.  References to technical details can be
found in the cited references.

We have already presented a didactic review of the global flow reconstruction
method (Varenna Lecture Notes, Buchler 1997) and there is no point in
repetition.  We therefore introduce the method slightly differently here in
\S2, and we emphasize the astrophysical results and applicability.

Can one always hope to extract information from a time-dependent signal about
the physical system that has generated this signal? The answer is clearly no.
First of all this would be hard if not impossible when the system is
nonautonomous (\ie the system itself changes on the time-scale of the signal).
Second, it would be impractical when the number of degrees of freedom is
neither large nor small.  In the extremes of a large number of degrees of
freedom (high dimensional phase-space) stochastic methods can be fruitfully
applied, whereas in the case of a small number of degrees of freedom (low
dimensional phase-space) the techniques from chaos theory apply.  This is this
latter type of system that we are concerned with here.

In \S 3 we briefly describe the R\"ossler system, a set of three first order
equations, which displays a behavior (Fig.~\ref{roessts}) that is still not
very commonly known or understood in the astronomical community, but is a
paradigm for certain types of irregular variable stars.  As we have seen this
system undergoes oscillations which appear almost mono-periodic in a Fourier
spectrum, yet undergo wild oscillations in amplitude.  An observational signal
with this Fourier spectrum would almost certainly be classified as periodic.

\section{Flow Reconstruction}

The basic idea that underlies the analysis that we are about to describe is
really extremely simple. We illustrate it with the help of an example that is
quite familiar to everybody, namely a second order differential equation with
constant coefficients, or equivalently a system of two first order ode's,
called a {\it two-dimensional flow}. 
\begin{equation}
 {dy\over dt} = f(y)
\label{defflow}
\end{equation}
where $y=(y_1,y_2)$.
We will omit most technical details and mathematical subtleties here.  Let us
sample the solution of this ode at equal time-intervals $\{t_i=t_\o + i\Delta
t\}$, and let us call these 'stroboscopic' values of the solution
$\{s_i=y(t_i)\}$. The solution is uniquely defined for all times provided we
know its values at {\sl two} anterior times.  Thus any $s_{i+1}$ can be
predicted from the knowledge of two preceding values, say, $s_{i}$ and
$s_{i-\Delta}$, where $\Delta$ is an integer greater or equal to 1.  Because we
are dealing with a given differential equation there exists a function $f$ such
that $s_{i+1}=f(s_i, s_{i-\Delta})$, no matter what the time index $i$ is.  The
form of this function depends of course on the differential equation and on the
size of the time-intervals $\Delta$.

It is convenient to introduce at this point the so called 'state vector'
 \begin{equation}
{\mathbf X_i} = (s_i, s_{i-\Delta})
\label{defX}
\end{equation}
and express the functional dependence with a map \F, viz.

\begin{equation}
 {\mathbf X_{i+1}} = {\mathbf F}({\mathbf X_i}).
\end{equation}
 The map \F moves us from a given state ${\mathbf X_i} = (s_i, s_{i-\Delta})$
to the next one ${\mathbf X_{i+1}} = (s_{i+1}, s_{i+1-\Delta})$ for all $i$.
The state vector \Xi maps out a stroboscopic representation of the trajectory
$y(t)$ in the two-dimensional 'phase-space' of the system.  (This is a
generalization of the usual physicist's definition of the phase-space
corresponding to position and velocity or momentum, \cf Weigend \& Gershenfeld
1994).  Because of the finite time-intervals the map {$\mathbf F$} is usually
nonlinear, even if the differential equation were linear.

We have chosen a dynamics that is described by a second order differential
system as an example only.  In fact it is oversimplified for our purposes
because irregular behavior can only occur for a flow in at least three
dimensions and only in the presence of nonlinearity (Ott 1993).  Generally
then, when the signal of interest is generated by a higher dimensional
dynamics, say a set of $d$ coupled first order ode's, we merely need to
increase the length or dimension of the vector to $d$

\begin{equation}
 {\mathbf X_i} = (s_i, s_{i-1},s_{i-2},\ldots,s_{i-d+1})
\end{equation}
 and the map ${\mathbf F}$ then operates in a $d$-dimensional phase-space,
while ${\mathbf X_{i}}$ describes the trajectory.

When confronted with an observational time-series we have no \apriori
knowledge, neither of the dimension $d$, nor of the form of the map \F.  We
shall call {\it embedding space} the space in which we try to reconstruct the
dynamics.  Suppose the embedding dimension $d_e$ that we have guessed is
actually smaller than $d$, then clearly we cannot suitably describe the
dynamical system.  In particular the embedded trajectory will exhibit
intersections which violates the uniqueness theorem of ode's.  On the other
hand there is nothing to be gained from making $d_e$ greater than
necessary. (This is very much akin to wanting to embed a 3-dimensional sphere
in a plane which obviously is not possible.  But, a 3D sphere remains a 3D
sphere even when we embed it unnecessarily in 4 or 5 dimensions.)  There exists
thus a optimal, minimal embedding dimension $d_e^{min}=d$ \th\th (There are
mathematical subtleties which sometimes can require $d_e^{min}>d$, but
$d_e^{min}<2d+1$\cf \eg Sauer \etal 1991.).  One of our goals is to determine
this $d_e^{min}$ which then also gives an upper limit of the physical dimension
$d$ of the unknown dynamics.  We shall also see below that another dimension,
the fractal dimension $d_L$, sets a lower limit to $d$.

\begin{equation}
 d_L \leq d \leq d_e^{min}
\end{equation}
 The two limits may coincide, as we have found in several examples, and in such
lucky situations, the procedure uniquely determines the dimension $d$ of the
physical phase-space.

For the nonlinear map ${\mathbf F}$ we make the simplest possible
choice (for which there are also good mathematical reasons),
namely we assume that ${\mathbf F}$ is a sum of all the
multivariate monomials up to some order $p$ that we can form with
the components of ${\mathbf X_i}$.

\begin{equation}
 {\mathbf F} = \sum_k {\mathbf{\beta_k}} {\cal M}_k({\mathbf X_i})
\end{equation}
 The ${\mathrm C^p_{p+d_e}}$ vector coefficients ${\mathbf \beta_k}$ of these
monomials can most easily be obtained by a linear least squares
approach with the help of the singular value decomposition method
(Serre, Kollath \& Buchler 1996a, hereafter SKB96).  Since the number of
fitting coefficients increases very rapidly with $p$ it is desirable to keep
the latter as low as possible.

 The flow reconstruction proceeds then as follows.  We successively assume
that the embedding dimension is $d_e=3, 4, \ldots$ and for each we determine
the function \F, \ie ${\mathbf{\beta_k}}$, by minimizing the error
\begin{equation}
 {\cal E} = \sum_i || {\mathbf X_{i+1}-F(X_i)} ||
\end{equation}
  where $||$ denotes the Euclidian norm, \ie the square root of the squares of
the components.  For ideal noise-free data the error levels off sharply at a
very small value (machine accuracy) when the minimum dimension $d^{min}_e$ is
obtained.  For noisy data the error levels off at some finite value and the
levelling-off is more gradual.

\begin{figure}
\centerline{\psfig{figure=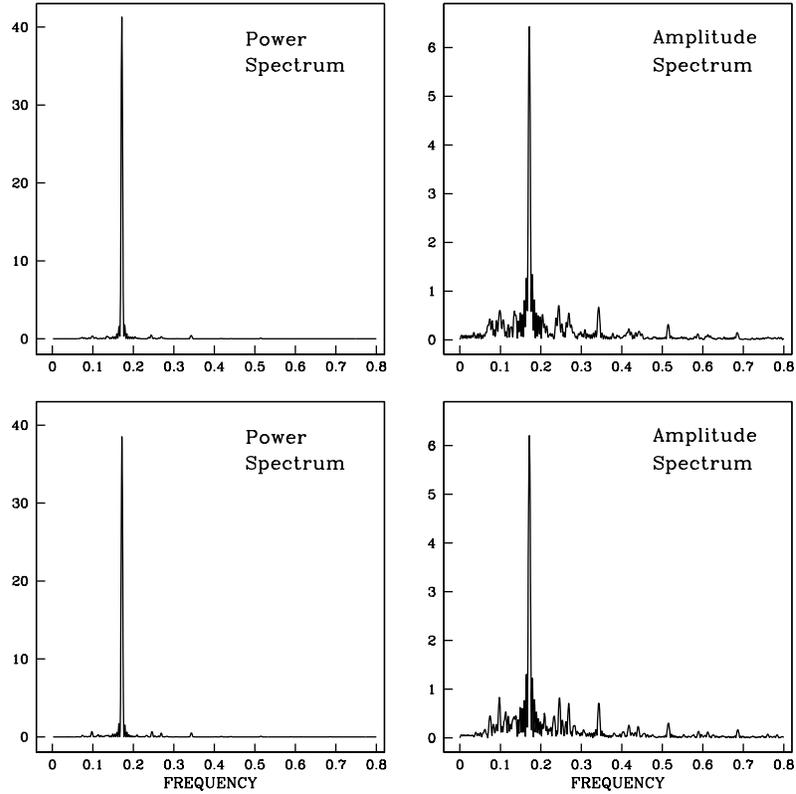,width=12cm}}
\caption{\noindent Fourier spectrum: Top: time-series (t=0, 200)
of Fig.~\ref{roessts}; Left: power, Right: amplitude; 
Bottom: Later segment of the same time-series (t=1848, 2048).}
\label{roessft}
\end{figure}

\section{The R\"ossler Oscillator}

Can we have confidence that in practice the global flow reconstruction approach
can really capture the dynamics of a system from the knowledge of the time
dependence of merely one of its variables?  Can we also be sure that it does
not erroneously find low dimensional chaos in any old irregular signal?

First we shall describe the behavior of the R\"ossler oscillator
and then we present the results of a series of analyses that we have made on
the time-series of the oscillator.

The R\"ossler system consists of an autonomous set of three
first order ode's (Thompson \& Stewart 1986).

\begin{eqnarray}
{dx_1\over dt} &\th =\th & -x_2-x_3 \\
{dx_2\over dt} &=& x_1+a x_2\\
{dx_3\over dt} &=& b +(x_1-c) x_3
\end{eqnarray}
 We have chosen the values $a=0.2$, $b=0.2$ and $c=4.8$ for the tests.

\begin{figure}
\vskip 10pt
\centerline{\psfig{figure=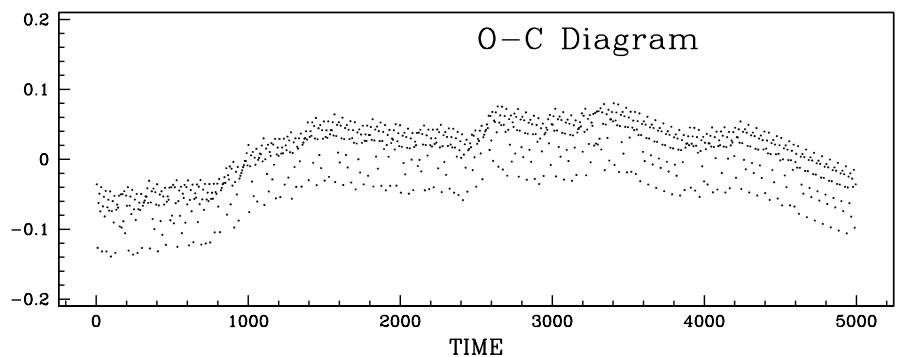,width=12cm}}
\caption{\noindent O-C Diagram of time-series of
Fig.~\ref{roessts}}
\label{roessomc}
\end{figure}

This is a deterministic system of equations, yet for these parameter values,
the solution of this system is irregular, \ie remains bounded, but never
repeats itself, and is very sensitive to initial conditions.  This means that
if we compute two solutions whose initial conditions differ by a tiny amount
there will always arrive a time when the two solutions diverge, \eg by
oscillating out of phase, no matter how small the initial difference.  In the
domain of flows (which are differential systems as opposed to maps which are
discrete) the R\"ossler equations are the simplest known paradigm for what is
called chaotic behavior.

In Fig.~\ref{roessts} displays a typical section of the temporal behavior of
the variable $x_1(t)$ that we obtain from integrating the R\"ossler equations.
One sees that there are wide fluctuations in amplitude.  Yet, one notices that
the phase is remarkably constant.  It is therefore no surprise that the Fourier
spectrum in Fig.~\ref{roessft} shows that the power is overwhelmingly in one
frequency $f_0=0.17$, plus a little power in the first few harmonics.
The more sensitive amplitude spectrum on the right shows some (unsteady) power
in a broad region centered on $f_0/2$ and $3f_0/2$.  It is unsteady in the
sense that different sections of a long time-series would produce different
structures.

Variable star observers have a predilection for O-C (observed minus calculated)
diagrams which are plots of the time of appearance of a feature (say the
light-curve maximum) relative to the corresponding multiples of the period.
The intended purpose of these plots is to uncover evolution which, when linear
in time, would show up as parabolic behavior. Because of its familiarity to
observers we show such an O--C diagram for a long time integration of $x_1(t)$.
It is clear that for an irregular signal such as the one we are discussing this
diagram adds nothing to one's understanding of the nature of the small but real
fluctuations in period. We recall that this system is {\sl not} evolving, since
the $a$, $b$ and $c$ parameters are constant!.  One notes a banded pattern of
clustering, for example, the higher concentration of points towards the top.
Its origin will become clearer when we see that there is a quasi
one-dimensional first return map that the dynamics follows \cite{roessretm},
similar to the well-know logistic map (Ott 1993).

It turns out that the familiar linear analysis is not very useful either.  To
illustrate this we have made a Fourier fit to the time-series $x_1(t)$ of
Fig.~\ref{roessts} (over 4000 time-units) with the twenty highest frequency
peaks from the spectrum of Fig.~\ref{roessft}. The resultant fit is shown as a
solid line in Fig.~\ref{roessfit} (left) together with $x_1(t)$ as a dashed
line. The multi-frequency Fourier fit is excellent, which is not astonishing
since one can obtain an arbitrarily good fit for a finite stretch of any
time-series as one increases the number of frequencies in the fit. However,
have we learned anything from such a fit about the time-series and how it was
generated? If we have then this fit should have 'predictive' power, \ie we
should be able to extrapolate it beyond the 4000 time-units and it should
continue to give a good fit.

 On the right side of Fig.~\ref{roessfit} we show a continuation of the
Fourier fit for the interval 20000 -- 24000 time-units, superposed on the
actual solution $x_1(t)$. The extrapolation deteriorates very quickly and at
this stage is already very different from the signal.  This shows that while it
is possible to obtain a very good interpolation of a complicated time-series
with a Fourier fit, the extrapolated signal bears little resemblance with the
physical signal.  Note that an ARMA process, which is also intrinsically
linear, does not fare any better (SKB96).  The obvious reason is that neither
the Fourier fit nor the ARMA process capture the nonlinear behavior of the
dynamics, and a more sophisticated technique is required.

In the observational astronomical literature one finds references to {\sl
nonlinearity} by which it is merely meant that the Fourier fit contains linear
combination frequencies of some more basic frequencies.  This could merely be
indicative of mode-coupling in a multi-periodic signal.  This 'nonlinearity'
has little to do with the intrinsically nonlinear nature of a chaotic
time-series that we are interested in here.

In contrast to the Fourier spectrum Fig.~\ref{roessretm} gives very important
information about the nature of the oscillations.  These diagrams are what is
known as 'first-return maps'.  They are simply a plots of $M_{k+1}$ versus
$M_k$, where the $\{ M_k\}$ are the successive maxima of the time-series (They
could also be the minima, or any other characteristic point of the signal).  On
the left side we show the first-return map for the $x_1(t)$ variable of the
R\"ossler oscillator.  The right hand side displays to the fits.  The thick
dots are the first return map of the multi-period Fourier fit, shown in
Fig.~\ref{roessomc} and the crosses the extrapolated fit.

The R\"ossler signal has a strong phase-coherence, and the first-return map has
a very typical quadratic shape that is typical of very low dimensional chaotic
systems (Ott 1993).

When we make a first-return map of the extrapolated Fourier fit we see that
extrapolated signal gives a scatter plot whose scatter increases with the
extrapolation time, showing again that while a multi-frequency Fourier fit
gives a decent interpolation, it does not capture any of the physics that is in
encrypted in the signal.

\begin{figure}
\centerline{\psfig{figure=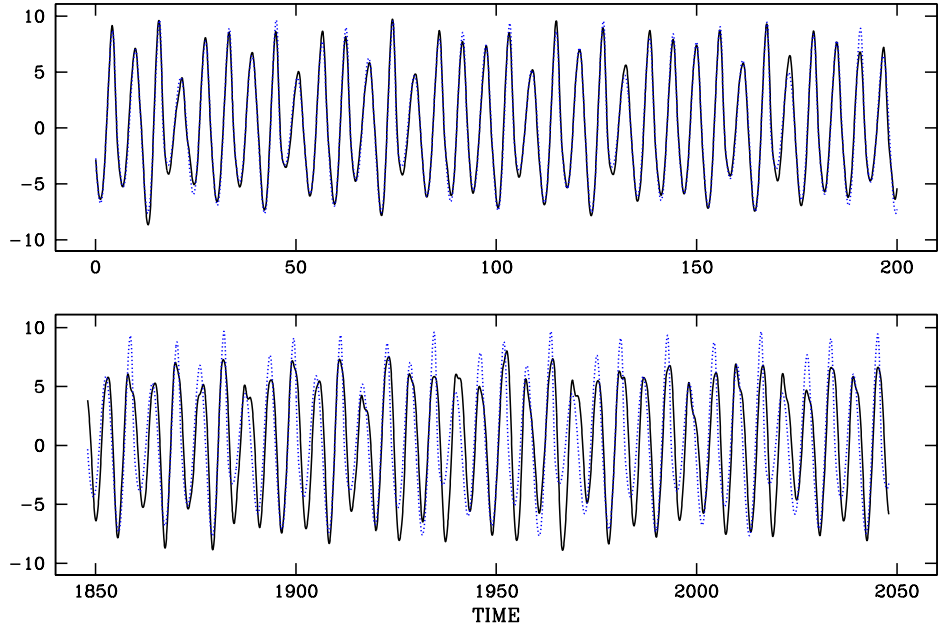,width=14cm}}
\caption{\noindent \small Fig.~\ref{roessretm}: 
Left: Fourier fit to  time-series of Fig.~\ref{roessts}
(t=0,200) ; right: Extrapolation of Fourier fit to t=1848, 2048.}
\label{roessfit}
\end{figure}

\begin{figure}
\centerline{\psfig{figure=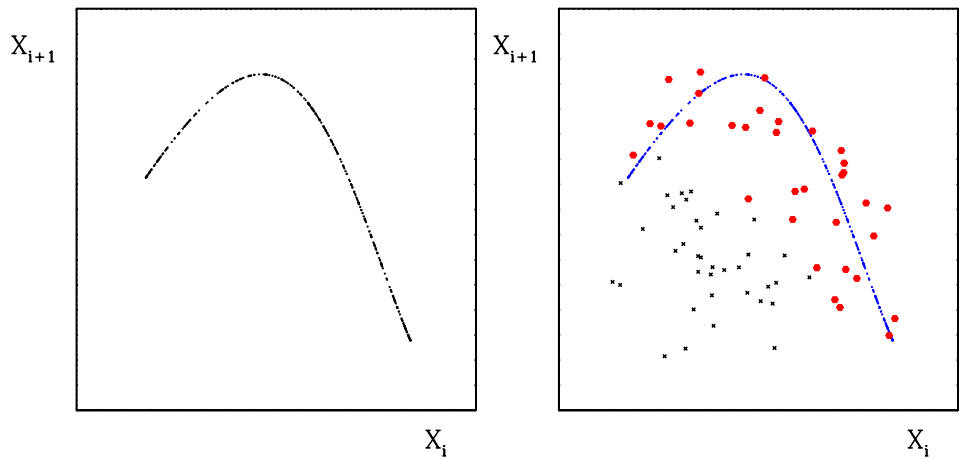,width=14cm}}
\caption{\noindent Left: First-return Map of the maxima of a long
stretch of the original R\"ossler time-series; right; first-return maps for the
interpolated (thick dots) and extrapolated (crosses) Fourier fit series,
superposed on the original return map.}
\label{roessretm}
\end{figure}

In SKB96 it is shown that the global flow reconstruction works very well.  From
$x_1(t)$ one can construct maps $M$ that capture the underlying R\"ossler
dynamics.  The success of the reconstruction manifests itself in several ways.
From the maps we can generate 'synthetic' signals trough iteration from a seed
value.  These synthetic signals have the same appearance as the training set
$x_1(t)$ that was used to generate the maps.  The maps and synthetic signals
are robust, \ie they are independent of some of the arbitrary parameters such
as the delay $\Delta$ or the maximum polynomial order $p$ over a comfortable
range of values.  Of course in the comparison it must be kept in mind that in
view of the chaotic nature of the signal is does not make any sense to compare
two time-series point by point.  Only average, global properties, matter.

In order to see whether the method might still, and perhaps erroneously,
predict low dimensional chaos SKB96 made the following test of the flow
reconstruction method using a stochastic signal with superficially very similar
properties as the R\"ossler time-series.  A test signal was constructed by
prewhitening $x_1(t)$ with the frequency 0.171 that dominates the Fourier
spectrum, as well as its first eight harmonics. The remaining signal (which is
quite erratic) is then Fourier analyzed, and we compute the envelope of the
complex amplitude spectrum and convolve its inverse Fourier transform with a
white noise signal.  The colored noise that we have thus generated is then
added to the original periodic pre-whitening signal to give our test signal.
The appearance of the signal is essentially the same as $x_1(t)$ and its
Fourier spectrum, by construction, is also very similar.

The important result of this test is that the flow reconstruction method was
not fooled into believing that this stochastic test time-series is a low
dimensional signal (SKB96).

Dynamicists have developed other characterizations of chaotic time-series and
of the underlying chaotic attractor.  Of particular interest to us are the
Lyapunov exponents which measure the degree of convergence of divergence of
neighboring trajectories (Ott 1993).  Unfortunately, observational data sets of
irregular or Semi-Regular variable stars are generally much too short to allow
us to compute Lyapunov exponents or a fractal dimension with any reliability.
However, once we have constructed the map or flow, we can use the latter to
compute the Lyapunov exponents, and from these the fractal dimension
\begin{equation}
d_L = K + {1\over|\lambda_{K+1}|} \sum_{i=1}^K \lambda_i
\end{equation}
 where the $\lambda_i$ are the Lyapunov exponents, ordered decreasingly, and
$K$ is the largest value such that the sum is positive.

If, as a test, we only take a short time-series of the R\"ossler oscillator we
could show that the Lyapunov dimension $d_L$ is reproduced quite accurately via
our flow reconstruction procedure, \eg one finds $d_L\approx 2.014$ instead of
the exact value of 2.013 (SKB96).

It might appear that we are getting a free lunch with this round-about way of
computing the Lyapunov dimension via the synthetic time-series. This is not so.
We have indeed used additional information, namely we have made the working
assumption that the signal is generated by a deterministic low dimensional
dynamics. As an illustration of this point suppose for example that we know
that we are given a temporal data set that has been generated by a second order
differential equation. From this we know that we can fit the time-series with a
map or flow that related only two successive points, and we can recover the
original differential equation and the general properties of its solution.

We have not yet talked much about the delay $\Delta$.  If we choose $\Delta t$
too small, the \Xi are very similar and the map \F becomes close to linear.
However, then the construction of \F is dominated by the noise.  On the other
hand, when $\Delta$ is too large, the function \F becomes very nonlinear and we
are obliged to go to high order monomials (large $p$) with a concomitantly
large number of coefficients to be determined. Clearly we have to choose an
intermediate regime, and the confidence in our reconstruction is enhanced when
this regime is sufficiently broad.

We note that in all the dynamics that we have reconstructed, namely the
test case of the R\"ossler oscillator (SKB96), the pulsations of a
hydrodynamic Pop. II Cepheid (W Vir) model (Serre \etal 1996b), the
observed light-curves of R Scuti (Buchler, Serre Koll\'ath \& Mattei 1995
[BSKM95];  Buchler, Koll\'ath, Serre \& Mattei [BKSM96])  and of AC
Herculis (Koll\'ath, Buchler, Serre \& Mattei 1998) we have found
that indeed the results are stable over a range of $\Delta$ values.

So far we have discussed only the search for a {\it map} that describes the
temporal evolution of the system in a stroboscopic sense.  It is also possible
to determine directly the differential system (flow) (Eq.~\ref{defflow}) that
describes the system system.  A direct determination of the flow is slightly
more delicate (SKB96).

\section{Variable Stars}

Variable stars are of course more complicated than the simple paradigm of the
3D R\"ossler system.  First, we are dealing with an infinite dimensional system
since a star has an infinity of modes of oscillation.  However, the relative
simplicity of dissipative systems as opposed to Hamiltonian ones, is that the
evolution does not occur the full infinite-dimensional phase-space, but only in
a subspace of finite, and often small, dimension.  Examples best known to the
reader are the mono-periodic pulsations of ordinary Pop I Cepheids and RR Lyrae
that occur in a mere 2D phase-space (in which the variable is the complex
amplitude of the excited mode), or the beat Cepheids, double-mode RR Lyrae and
bump Cepheids that occur in a 4D phase-space because here two pulsation modes
are involved.  One hopes, and one indeed finds, that the same 'dimensional
reduction' remains true for the more complicated, and highly irregular
pulsations of W~Vir and RV~Tau stars.

There is another difference at a more subtle level.  In the previous section,
when we studied the dynamics of the R\"ossler oscillator we used one of the
phase-space variables, namely $x_1(t)$, in the reconstruction.  In the stellar
case we have at our disposal only the temporal behavior of the luminosity which
is not a phase-space variable.  It can be shown however (\eg Abarbanel \etal
1993, Weigend \& Gershenfeld 1993) that even so we can infer the properties of
the dynamics provided that the luminosity contains all the information about
the dynamics, in other words provided that $L$ is a generic function of the
phase-space variables.  The strong coupling between the heat flux and the
pulsation variables (radii and velocities) guarantees that the luminosity is
indeed a 'good' variable.

When dealing with observational data two further practical complications
arise. First, observations are never made at equally spaced time-intervals and
we need to interpolate, which introduces errors or noise.  Second, there is
also observational noise, and if the noise level is too high it may be very
difficult or impossible to extract the dynamics from the time-series.

Very few observational data sets in which one may expect low dimensional chaos
are sufficiently long and sufficiently densely sampled to perform a flow
reconstruction.  Fortunately the AAVSO has archived amateur astronomer
observations of variable stars over the years. There are two such stars,
R~Scuti and AC~Herculis, belonging to the RV~Tau type of Pop II Cepheids that
undergo irregular pulsations of large amplitude and are very bright, and thus
have been observed very assiduously by amateur astronomers.  Because the
observations are mainly visual, the observational errors are relatively large,
$\sim$ 0.1 mag, but because there are a lot of statistically independent data
we can still extract a reasonably accurate light-curve.

In Fig.~\ref{aavso} we display small sections of the observational data of
R~Sct and AC~Her superposed on our spline-fit that we use for the
reconstruction.  The observational error is the same for the two stars, but the
amplitude is smaller for AC~Her so that the latter has a worse signal-to-noise
ratio than AC~Her.

\begin{figure}
\centerline{\psfig{figure=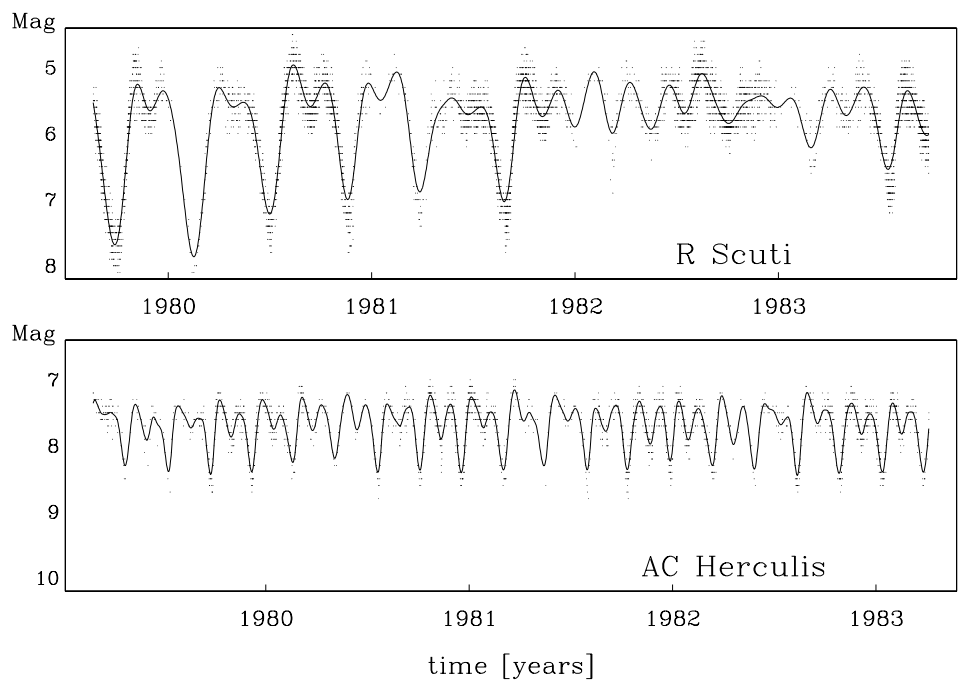,width=13cm}}
\caption{Segments of the AAVSO light-curve data and our spline-fits.}
\label{aavso}
\end{figure}

\begin{figure}
\centerline{\psfig{figure=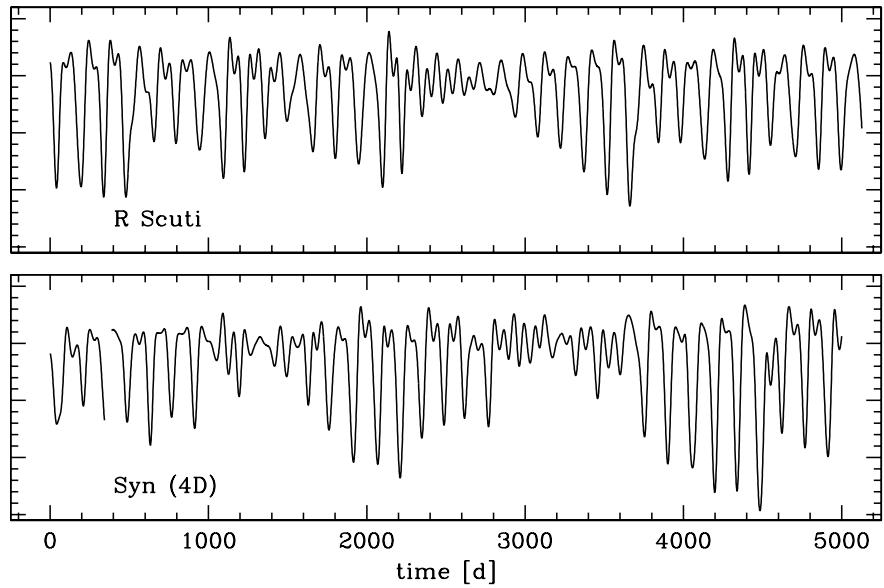,width=12cm}}
\caption{Top: R~Sct light-curve; Bottom: synthetic fit (see text)}
\label{rsctmag}
\end{figure}

\begin{figure}
\centerline{\psfig{figure=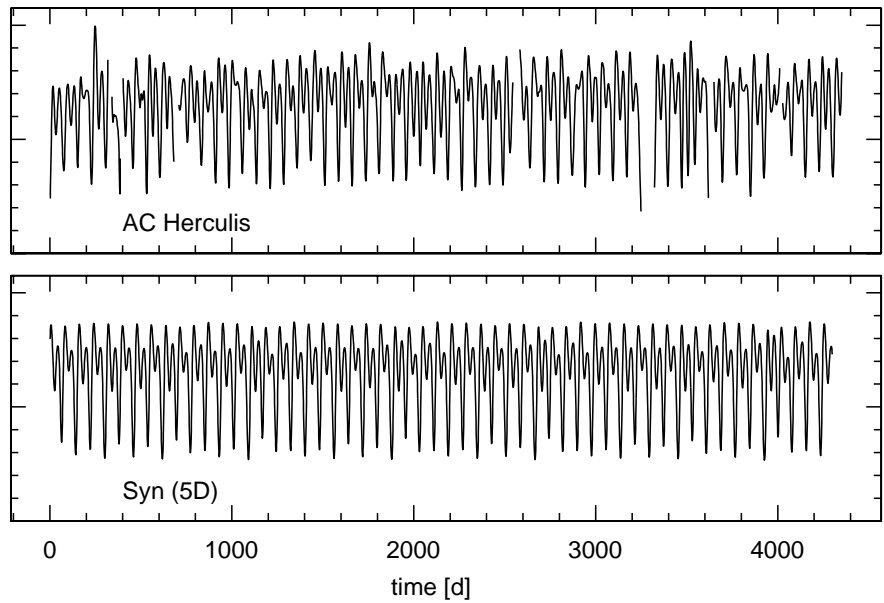,width=12cm}}
\caption{Top: AC~Her light-curve; Bottom: synthetic fit (see text)}
\label{achermag}
\end{figure}

The complete time-series that we have utilized for our flow reconstruction are
shown in the top sections of Fig.~\ref{rsctmag} for R~Sct, and in
Fig.~\ref{achermag} for AC~Her.  The corresponding amplitude Fourier spectra
are displayed in Fig.~\ref{rsctft}.

\subsection*{R Scuti}

In a Fourier analysis of 150 years of R~Sct light-curve data it was shown by
Koll\'ath (1990) that the Fourier spectrum has a more or less constant envelope
with large peaks in the broad vicinities of $\approx 0.07 d^{-1}$ and its
harmonic.  But the spectrum is not steady since for successive segments of the
data the individual peaks occur in different places and without any smooth
transition between them.  This rules out the light-curve as being that of an
evolving multi-periodic star.  Consequently, while multi-periodic fits might
provide good interpolations of the light-curve, they are useless from a
physical point of view because they add no new understanding of the mechanism
for the irregular behavior.

\begin{figure}
\centerline{\psfig{figure=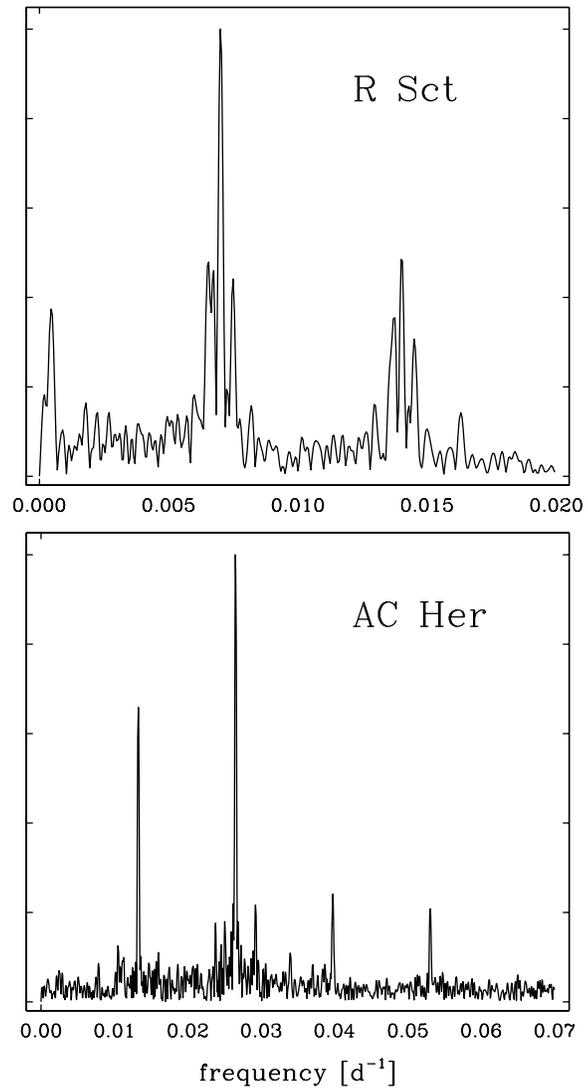,width=8cm}}
\caption{Amplitude Fourier spectra of R~Sct and AC~Her.}
\label{rsctft}
\end{figure}

In BSKM95 and BKSM96 it was further demonstrated that an ARMA
process cannot underlie the light-curve.  Again, it is not astonishing that
such a linear analysis does not work either because they are not independent.
Fortunately, an ARMA process does not capture the dynamics, because if it did
one would be hard pressed to come up the a stochastic mechanism that can give
rise to factors of 40 fluctuations in luminosity!

There is no point in repeating the detailed analysis of R~Sct that has been
described in BKSM96.  That paper followed the procedure outlined in \S3.
Starting with dimension $d_e>2$, maps were constructed in spaces of increasing
dimension and for a range of values of time-delay $\Delta$ and maximum
polynomial order $p$.  Once one has obtained such maps, one can easily iterate
them starting with a seed value and thus generate synthetic signals.  One of
the criteria for claiming success is that such synthetic signals have the same
properties, in particular the same appearance, as the stellar data that were
used to generate the maps.  The resemblance of these synthetic signals with the
observational data is an indication that our reconstructed map has captured the
dynamics, but additional tests have to be made to validate the reconstruction.

For R~Sct all attempts to obtain faithful maps failed in 3D.  But BKSM96 found
that it was possible to obtain maps that show robustness within some range of
delays $\Delta$ and polynomial nonlinearity $p$ for embedding dimensions
greater of equal to four.  The minimum embedding dimension for R~Sct is
therefore $d_e^{min}=4$.

Because of the shortness of the observational time-series it is not possible to
derive Lyapunov exponents directly from the data, but these exponents can be
derived from the maps or from the synthetic signals.  BKSM96 found that the
Lyapunov exponents also show good robustness, and that they indicate a fractal
dimension of $d_L\approx 3.1$.  What is important is that this dimension is
essentially invariant even when the embedding dimension $d_e$ is as large as 6.

The fractal dimension of the attractor imposes a lower bound of $d_L \leq d$ on
the dimension of phase-space, whereas the minimum embedding dimension $d_e\leq
d $ imposes an upper bound of 4.  In the case of R~Sct the upper and lower
bounds agree and we can conclude that the dimension of the physical phase-space
is $d=4$.

What is the physical meaning of this dimension?  This becomes clear when we
look at the linear eigenvalues of the map about the fixed-point, which
correspond to the linear vibrational stability of the star.  We have displayed
these quantities for our best map in Table~\ref{table1}.  The stability roots
are written in the form $\exp (\rho +i\nu)$.  Negative and positive $\rho$
therefore denote linear stability and instability, respectively.  The (angular)
frequency eigenvalue, $\nu_1=0.0068$ is close to the average cycling frequency
of the light-curve, whereas the higher frequency is almost double its value.
The lower frequency is unstable and the higher frequency is stable.  Because we
can associate these eigenvalues with vibrational modes this then leads to the
nice and simple physical picture: The 4D phase-space that we have uncovered in
the light-curve of R~Sct is spanned by two vibrational modes, a linearly
unstable mode and a resonant, linearly stable overtone.  One can therefore
interpret the irregular pulsation as a continual exchange of energy between two
pulsation modes, one that wants to grow and one, with approximately twice the
frequency, that wants to decay.

Why do Pop II Cepheids such as R~Sct behave so differently from their Pop I
siblings?  The reason is that they have much larger relative growth-rates
$\eta=2\kappa P$.  In the classical, Pop I Cepheids the growth-rates of the
excited modes are generally of the order of a percent or less.  This means that
amplitude variations can only occur on time-scales of hundreds of pulsations,
whereas the relative growth-rates of the Pop II Cepheids are of the order of
0.5 or greater.  Obviously irregular behavior can only occur when the amplitude
can change on a time-scale of a period.  The reason that Pop II Cepheids have
higher relative growth-rates is due to their higher luminosity/mass ratio --
the coupling between the radiative flux and the pulsation is therefore much
stronger.  In the jargon of stellar pulsation theory, these stars are much more
nonadiabatic.

\begin{table}[h]
\caption{Linear stability roots of fixed point of the R~Scuti maps}
\begin{tabular*}{\hsize}{@{\extracolsep{\fill}}cccccc}
\hline
$\hf \nu_1$ \hf & $\hf \rho_1$ \hf & ~ &
$\hf \nu_2$ \hf & $\hf \rho_2$ \hf & \th \cr
\hline
  0.0068 &  0.0044 & & 0.0145 & --0.0062 \cr
\hline
\end{tabular*}
\label{table1}
\end{table}

One final comment.  One might object that it may be possible to generate a map,
but that there does not necessarily exist any flow of which the map is the
stroboscopic manifestation.  In our analysis of R Scuti we have used delays
$\Delta$ corresponding to a day (compared to the cycling time of 70 days).
Because of these fine time-steps the map \F is very close to a (differential)
flow.  We find that one of the Lyapunov exponents is always close to zero (For
a flow it should be rigorously zero.)  This further corroborates our belief
that we have captured the dynamics of the star.

\subsection*{AC Herculis}

The analysis is more delicate for the AC~Her data which, as we have
already pointed out, have a smaller sign-to-noise ratio.  Details of the
reconstruction are described in Koll\'ath, Buchler, Serre \& Mattei (1998).
The best map one can obtain is not as satisfactory as for R~Sct.  A synthetic
light-curve is shown in the bottom of Fig.~\ref{achermag}.  The fractal
dimension turns out to be a little smaller, $d_L\approx 2.3$, presumably
because of the more regular behavior of the light-curve compared to R~Sct.

Before finishing this section we want to point out that the same flow
reconstruction method has also been applied to the solar cycle problem by Serre
and Nesme-Ribes (1997, 2000).

\subsection*{Data requirements}

An observer may want to know what the best strategy might be for observing an
irregular variable star.  This is a difficult question to answer in a general
sense, but some general comments are possible (\cf SKB96).  We recall that we
want to be able to construct as many as possible \Xi vectors with the least
amount of noise, therefore

(a) the signal-to-noise ratio has to be large enough so that the stochastic
behavior does not overwhelm the deterministic one;

(b) it is important to have very good coverage -- this may be 20 or more points
per cycle.
If we have to interpolate over small gaps this is equivalent to introducing
noise which destroys our ability to extract the dynamics; larger gaps are ok as
long as the star (and its dynamics) does not change during the gap.

(c) the time-series must be typical of the dynamics.  By that we mean that it
must explore essentially all of phase-space.  Common sense tells us that if,
for example, the signal has long term modulations, then it is necessary to take
a stretch of data that is long enough to sample those modulations.

\section{Time-Frequency Analysis}

In this section we discuss the application of modern time-frequency analysis to
variable star data, even though it is a {\it strictly linear analysis}.  The
reason is that one can gain a useful overview of how the signal changes in
time, and, as we see, it confirms our conjecture about R~Scuti's pulsations as
being due to the continual exchange between two modes of pulsation.  We stress
though that this approach, as all other linear ones, is purely interpolative,
has little if no extrapolative potential, and therefore does not allow us to
gain information about the nonlinear physical nature of the source of the
signal.

\subsection*{The Transforms}

The Fourier transformation of $s(t)$ 
$$S(\omega) = {\cal F}_{t, \omega} \left [s(t)\right ]
 = \int s(t)\exp(-i\omega t) d t$$
is widely used to determine the spectral content of stationary signals.
However for time-series with variable frequency content the Fourier
spectrum can tell nothing about the temporal variation.
Note that we use the
above notation for the Fourier transformation herein, and
${\cal F}_{t,\omega}^{-1} \left [S(\omega)\right ]
 = 1/(2\pi)\int S(\omega)\exp(i\omega t) d t$ for the inverse Fourier
transformation.

A variety of techniques that are extensions of Fourier analysis to nonsteady
time-series, such as G\'abor transforms and wavelet analysis, and more
sophisticated time-frequency distributions, have made it into the astronomical
literature over the last decade.  A nice general reference to time-frequency
analysis is Cohen (1994).  Several applications to astronomical data have been
summarized in recent workshop proceedings (\eg Buchler \& Kandrup 1997).

The spectrogram \ie the square of the short time Fourier transform (STFT)
is a classical tool of time Frequency analysis. The general form of STFT
is
given by:
\begin{equation} G(t,\omega) =
 {\cal F}_{\tau,\omega} \left [ s(\tau) h^*(\tau-t) \right ],
\label{gabor}
\end{equation}
where the kernel $h(t)$ performs a local (in time)  weighting on the
signal. The STFT introduced by G\'abor (1946), who used a Gaussian kernel:
\begin{equation}
 h(t) = \exp(-t^2/(2\sigma^2)).
\end{equation}
We refer to this special STFT as the G\'abor transform.

Wavelets are also frequently applied for astronomical data sets, although they
do not perform as well as the other methods.  Here we note only that replacing
the constant $\sigma$ in the G\'abor kernel (Eq.~\ref{gabor}) by
$\sigma(\omega) = \omega / c$ results in the Morlet wavelet.  With this
connection a numerical realization of the G\'abor transform can be easily
changed to wavelet and vice verse.

The simplest member of a more general class of time-frequency
distributions (see Cohen 1966) is the Wigner distribution (WD)
(Wigner 1932):
\begin{eqnarray}
  W(t,\omega) &=& \int \exp(-i\tau\omega)
  s^*(t - {\tau\over 2}) s(t + {\tau\over 2}) d\tau \\
  &=& {\cal F}_{\tau,\omega}
  \left [s^*(t - {\tau\over 2}) s(t + {\tau\over2}) \right ]
  \label{wd}
\end{eqnarray}
The WD has numerous advantages, like the properties of
marginals, it recovers the nature of a chirp signal exactly etc. However,
because of the nonlinearity of the transformation, the WD of
multicomponent signals is contaminated by cross terms (see Cohen 1989).

The advantageous properties of the WD can be further expanded 
with more flexibility in the Cohen's class of TFD's (Cohen 1966):
\begin{equation}
  C(t,\omega) = {1\over 2\pi} \int\int\int
  \exp(-i\xi t - i\tau\omega + i\xi \theta) \Phi(\xi,\tau)
  s^*(\theta - {\tau\over 2}) s(\theta + {\tau\over 2})d\theta d\tau d\xi,
\label{gtfd}
\end{equation}
where $\Phi(\xi,\tau)$ is the kernel of the distribution.
The Wigner Distribution  is a special case  by the unit kernel
($\Phi(\xi,\tau)=1$).

Alternatively $C$ can be expressed in the terms of the WD according to:
\begin{equation}
  C(t,\omega) = {1\over 2\pi} \int\int
  \varphi(t-\theta,\omega-\xi) W(\theta,\xi) d\theta d\xi.
\label{gtfd2}
\end{equation}
Here $\varphi$ is the 2D Fourier transform of the kernel $\Phi$:
\begin{equation}
\varphi(t,\omega) =  {\cal F}^{-1}_{t,\xi}
\left [ {\cal F}_{\tau,\omega}
\left [ \Phi(\xi, \tau) \right ] \right ] .
\end{equation}
 If $\varphi$ is a low-pass filter in the time-frequency plane, then the TFD is
a smoothed version of the WD. The smoothing compresses the cross terms, but
decreases the resolution in the distribution.

For the numerical realization it helps to calculate the TFD's through
Fourier transformations. It is easy to see that Eq.~\ref{gtfd} can be
written in the following form:
\begin{equation}
  C(t,\omega) = {\cal F}_{\tau,\omega} \left [
  \int \hat\varphi(t-\theta,\tau)
  s^*(\theta - {\tau\over 2}) s(\theta + {\tau\over 2})d\theta \right ],
\label{gtfd3}
\end{equation}
where 
$\hat\varphi(t,\tau) ={\cal F}_{t,\omega} \left [ \Phi(\omega,\tau)\right ].$
The TFD is the Fourier transform of the local autocorrelation
\begin{equation}
R_t(\tau) = \int \hat\varphi(t-\theta,\tau)
s^*(\theta - {\tau\over 2}) s(\theta + {\tau\over 2})d\theta
\end{equation}
For our applications we use the TDF kernel defined by Choi and Williams
(1989): 
$\Phi(\xi,\tau) = \exp(-\theta^2\tau^2/\sigma)$,
giving the following distribution. Then in Eq.~\ref{gtfd3} the weighting
function for the local autocorrelation becomes:
\begin{equation}
  \hat\varphi(t,\tau) =
  {1\over 2\pi^{1/2}}   \left ( {\sigma \over \tau^2}\right )^{1/2}
  \exp\left (- {t^2\over 4} {\sigma \over \tau^2} \right ).
\end{equation}

We note that the separable exponential kernel: 
\begin{equation}
\Phi(\xi,\tau) = \exp(-\tau^2/\alpha^2 -\beta^2 \xi^2)
\end{equation}
 is also very effective in
suppressing the cross terms, but contrary to the Choi-Williams
distribution a TFD with this kernel does not satisfy the marginals.
The separable exponential kernel has a special property, \eg
with  $\alpha = \beta$ it reproduces the G\'abor transform, while
with $\beta << \alpha$ the distribution converges to the WD.
This makes it possible to get a continuous set of TFD's connecting
the Wigner distribution to the STFT.

One cannot claim \apriori that any one of these methods is generally
superior to the others.  This depends largely on the nature of the signal
and of what features one tries to enhance, and it is therefore
advantageous to use simultaneously several of them.

\begin{figure}[ht]
\centerline{\psfig{figure=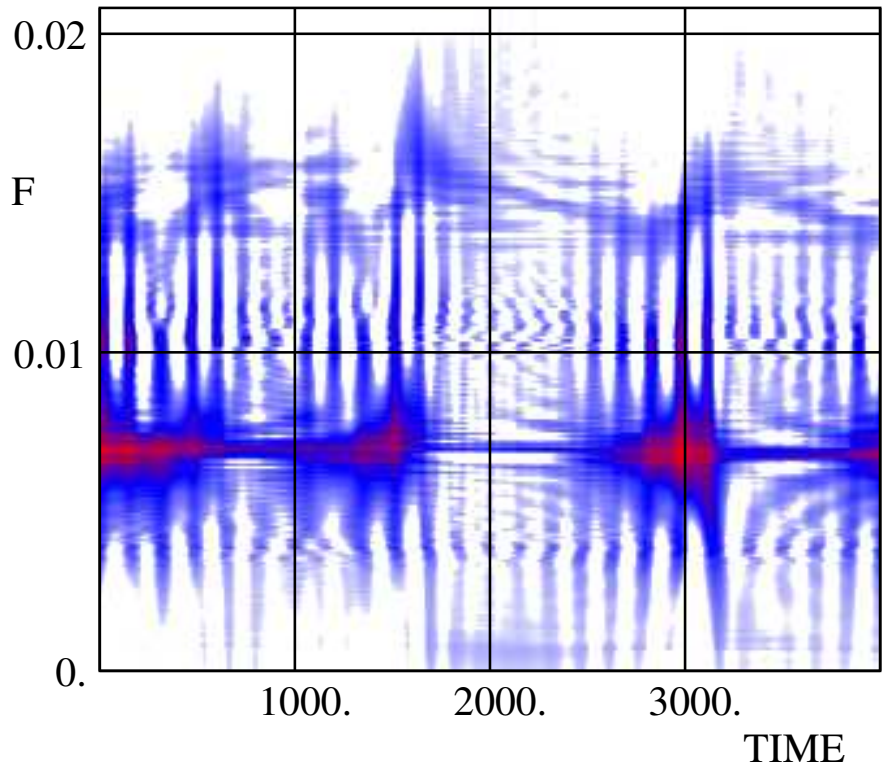,width=12cm}}
\caption{Choi Williams distribution of the synthetic R Sct light-curve}
\label{scwd}
\end{figure}

\begin{figure}[ht]
\centerline{\psfig{figure=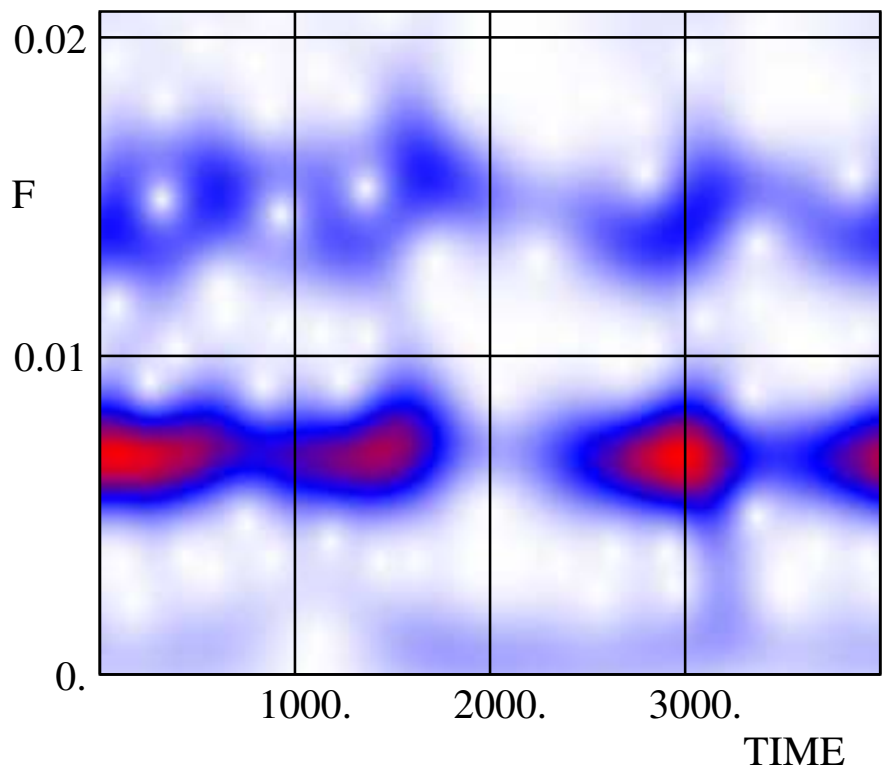,width=12cm}}
\caption{G\'abor transform of the synthetic R Sct light-curve}
\label{sgab}
\end{figure}

\subsection*{Application}

We demonstrate the power of time frequency distributions on a synthetic signal
obtained by the iteration of the map fitted to the R Sct observations (BKSM96).
We selected a 4000 day long data segment (top curve in Fig.~\ref{sdec2}) of a
synthetic R~Sct light-curve for the time frequency analysis. As we have already
shown in Table~\ref{table1} the linear stability analysis of the fixed point of
this map uncovers a pair of frequencies $f_\o\approx 0.007$ c/d and $f_1\approx
0.016$ c/d. The interaction of these two spiraling manifolds is very important
in the dynamics of the map.  The above mentioned frequencies and $2f_\o$
manifest themselves in the Fourier spectrum but with broad structures of peaks.

Fig~\ref{scwd} shows the Choi-Williams distribution of the synthetic data. The
amplitude modulation of the different frequency components 
and the shift between the frequencies $2f_\o$ and $f_1$ are clearly
visible in the time-frequency representation of the signal. This
effect is the result of the interaction of the two spiral manifolds.

For comparison we have also shown the G\'abor transform for the same signal in
Fig.~\ref{sgab}.  Although the same overall features are the same in the
G\'abor transform the latter appears like an out-of-focus version of the much
sharper Choi-Wiliams distribution.

In the study of multi-component signals with strong modulations it is usually
fruitful to decompose the signal for its physically relevant components. The
analytic signal formalism (see Koll\'ath and Buchler in this Volume) can be
used to obtain the instantaneous amplitudes and periods of the different
components.  In Figure~\ref{sdec1} we present the real part of the filtered
analytic signal for the three main components around the frequencies $f_\o$,
$2f_\o$ and $f_1$. The half width of the filtering was set to 0.0015 c/d. The
amplitudes of $f_\o$ and its harmonic $2f_\o$ are correlated as expected.  The
two higher frequency components cannot be exactly separated because of the
small frequency difference, but the most important tendencies are visible. The
amplitude of the $f_1$ component starts to increase at the maximum amplitude of
the lower frequency part.

\begin{figure}[ht]
\centerline{\psfig{figure=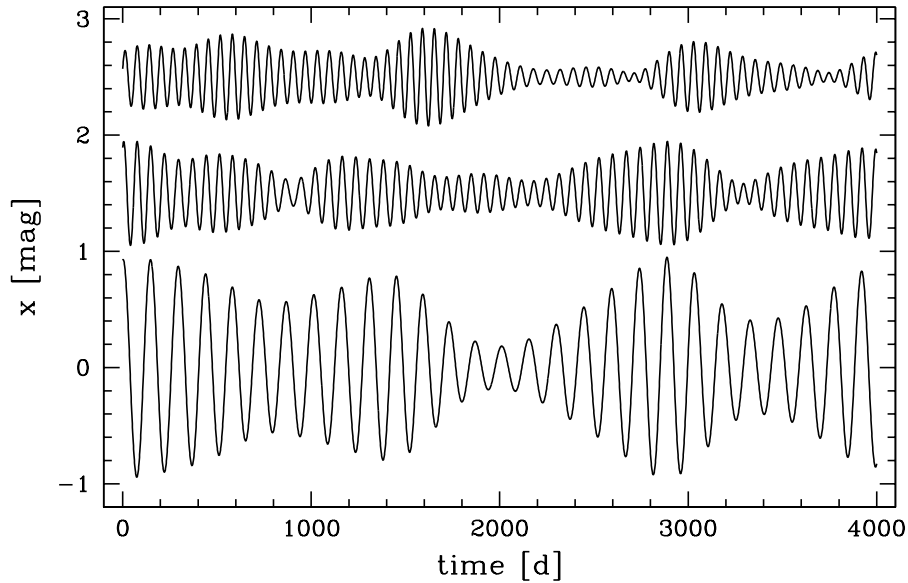,width=12cm}}
\caption{The decomposition of the signal presented in
Fig.~\ref{sdec2}: lower curve: $f_\o$, middle: $2f_\o$,
top: $f_1$.}
\label{sdec1}
\end{figure}

To check the quality of the decomposition, the sum of the components together
with the original synthetic variation is presented on Figure~\ref{sdec2}.
Although some of the local features are missed in the three component
representation, it gives a nice agreement with the data. The amplitude is
slightly decreased because of the windowing in Fourier filtering.

\begin{figure}[ht]
\centerline{\psfig{figure=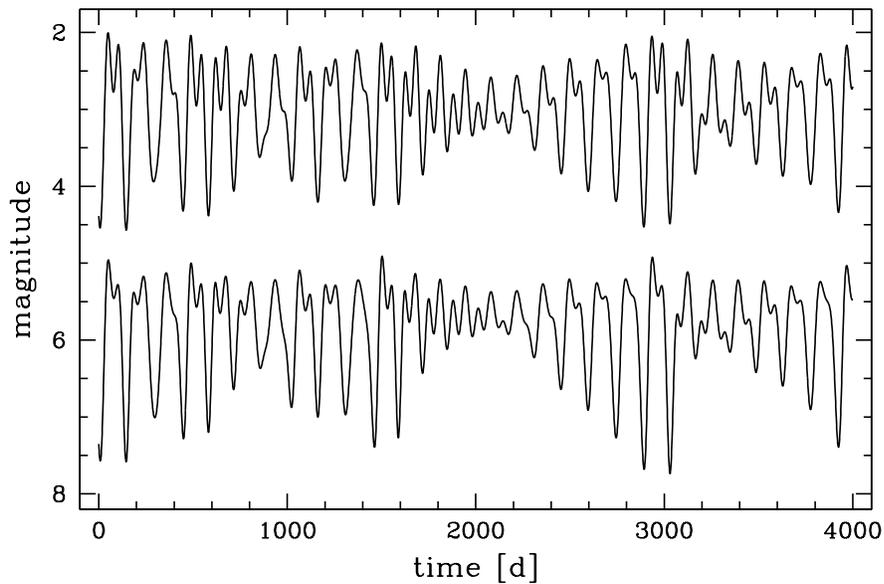,width=12cm}} \caption{The
synthetic light-curve from the R Sct map (top) and the sum of its
three primary components (bottom).} \label{sdec2}
\end{figure}

For the R Sct light-curve itself the frequency separation of the higher
frequency components appear somehow smaller, and it is difficult to accurately
decompose those parts, but the various TFD's still can recover the variable
frequency content around f=0.015 c/d.  In an application to the light-curve of
the star Koll\'ath \& Buchler (1997) found that the standard wavelet analysis
gave very little insight, but that the G\'abor transform and the Choi-Williams
distribution showed a systematic wavering of power in the neighborhood of twice
the basic frequency.  During the phase when the pulsation amplitude of the
linearly unstable basic pulsational mode of frequency $f_\o$ grows, the
amplitude of its first harmonic (2$f_\o$) grows concomitantly, but during the
phase when the linearly stable, but resonant overtone mode with a frequency
$\approxgt 2f_\o$ dominates (and the verall amplitude decays) we see the
frequency power shift from $2f_\o$ to a frequency $\approxgt 2f_\o$.  This
picture supports the basic properties of the map that were obtained with the
flow reconstruction method described in the previous section.

We end this section by stressing again that the time-frequency analysis, being
linear, it cannot capture the nonlinearity of the dynamics.

\section{Topological Analysis}

The topological analysis is much more mathematical and its
description is beyond the scope of this review.  We therefore only
give the gist of it here and we refer to Letellier \& Gouesbet
(1997) and Letellier \etal (1996) for a detailed description and
its application to variable stars.  The topological analysis can
be very useful in characterizing and comparing different
time-series and their underlying attractors, but it is limited to
three-dimensional dynamical systems and in addition to those that
have a large intrinsic disspation.

Chaotic attractors can be regarded as the set of all their
unstable orbits (Ott 1993).  If these orbits are not too unstable,
\ie the system stays in such an unstable orbit for a good fraction
of a period and if a sufficient number of visits occur in the span
of the time-series then we are able to extract enough information
about these orbits to characterize them.  The way these orbits are
braided in phase-space gives us a characteristic fingerprint of
the attractor (and thence the dynamics).  The method extracts
information about the lowest few unstable orbits of the attractor,
sets up a template for the attractor, and then checks that the
higher unstable orbits appear in the order predicted by the
template.

The restriction of the analysis to 3D comes from the fact that the analysis
makes use of knot theory (for wich there is no theory in 4D).  On the other
hand, the restriction to large dissipation is required if we are to be able
characterize the organization of the attractor into bands.  This requires then
that that the fractal dimension of the attractor be 2+$\epsilon$, with
$\epsilon\approxlt 0.1$.

So far the method has been successfully tested on tried on the R\"ossler
oscillator and applied to the nonlinear pulsations of models of W Virginis
stars.  It turns out that the dynamics of the W Virginis model is very similar
to the R\"ossler dynamics, although the templates of the dynamics are slightly
different (\cf  Letellier \& Gouesbet 1996) and Letellier \etal 1997)

The fractal dimensions of the attractor of R Scuti ($\sim$ 3.1) and of AC
Herculis ($\sim$ 2.3) are unfortunately too large for this method to work
though.  This lack of applicability can also be inferred from the shapes of the
first-return maps of these pulsating stars (Buchler, Koll\'ath \& Serre 1995,
Figs.~4 and 5) which are certainly not 1D.  The topological analysis relies on
the existence of a fairly well defined first-return map in order to define a
symbolic dynamics and the pruning tree (nature and order appearance of the
higher order unstable orbits).

\section{Conclusions}

We have discussed how the standard linear techniques of time-series analysis
become useless when the aim is to understand the nature of an intrinsically
nonlinear time-series.  We have presented two recently developed tools, the
flow reconstruction method and the topological method, both of which take the
nonlinear nature of the signal into account.

The irregular variability of the Semi-Regular variables had remained a mystery
for a long time.  With the help of the flow reconstruction approach it has now
been solidly established that this irregular variability is the result of a
chaotic dynamics that governs the behavior of these stars. The approach has
also laid bare the physical mechanism: the irregular variability comes from the
nonlinear and resonant interaction of just two radial pulsation modes, one
linearly unstable and the other linearly stable, but with approximately twice
the frequency of the first.  It is the highly nonadiabatic nature of these
modes that allows amplitude variations to occur on the time-scale of a
'period'.  Future research will have to show how and why the pulsations of the
Pop II Cepheids, comprising the W~Virginis and RV~Tauri type stars, become
increasingly complicated as their luminosity increases.  A good quantitative
understanding of this behavior would generate a tool for nonlinear
asteroseismolgy.

The topological analysis has been applied to the nonlinear pulsations of a
model for a W~Virginis type star, confirming the chaotic nature of the
pulsations and giving a quantitative comparison with the R\"ossler system.  The
method will be very useful for the analysis of the light-curves of W~Virginis
type stars when they become available.  The pulsations of their more luminous
siblings, the RV ~Tauri type stars are unfortunately too high dimensional for
this method.

Finally, we have discussed a time-frequency method.  This linear method is
purely descriptive, but it is useful for shedding light on the nature of
unsteady pulsations in that it shows the temporal variations of the modal
content of the pulsations.

 \vskip 2cm

\begin{acknowledgments}

 It is a pleasure to acknowledge valuable discussions with our collaborators
Thierry Serre, Gerard Gouesbet, Christophe Letellier and Janet Mattei. This
work has been supported by the National Science Foundation (AST95-28338,
AST98-19608) and by the Hungarian OTKA (number T-026031).

\end{acknowledgments}

\end{document}